\documentstyle[preprint,aps]{revtex}

\newcommand{\be}{\begin{eqnarray}}
\newcommand{\ee}{\end{eqnarray}}
\newcommand{\BE} {\begin{equation}}
\newcommand{\EE} {\end{equation}}

\begin{document}
\draft

\title{
\begin{flushright}
{\normalsize IP-ASTP-27-94 \\
\vspace{-0.2in}
NHCU-HP-94-32}
\end{flushright}
\Large\bf Heavy Quark Effective Theory on the Light-Front}

\author{{\bf Wei-Min Zhang,}$^a$\thanks{e-mail address:
                wzhang@phys.sinica.edu.tw}
        {\bf Guey-Lin Lin,}$^b$\thanks{e-mail address:
                glin@bunun.iop.nctu.edu.tw}
        {\bf and Chi-Yee Cheung}$^a$\thanks{e-mail address:
                phcheung@ccvax.sinica.edu.tw} \\
        $^a$Institute of Physics, Academia Sinica, Taipei 11529, Taiwan\\
        $^b$Institute of Physics, National Chiao-Tung University, Hsinchu
                30050, Taiwan\\ }

\date{Dec. 25, 1994}

\maketitle

\begin{abstract}
The light-front heavy quark effective theory is
derived to all orders in $1/m_Q$.
In the limit $m_Q\rightarrow \infty$, the theory exhibits
the familiar heavy quark spin-flavor symmetry.
This new formalism permits a straightforward
canonical quantization to all orders in $1/m_Q$;
moreover, higher order terms have rather simple
operator structures.  The light-front heavy quark effective theory
can serve as an useful framework for the study of non-perturbative QCD
dynamics of heavy hadron bound states.
\end{abstract}

\newpage

\baselineskip .29in

\noindent {\bf 1. Introduction}

Since its discovery a few years ago\cite{Isgur90},
the concept of heavy quark spin-flavor symmetry (HQS) has become
a very useful tool for analyzing transition processes involving
heavy hadrons\cite{review}.
The HQS is most useful in relating otherwise unrelated transition
form factors of heavy hadrons, thus greatly simplifying the
complexity of theoretical analyses.
For instance, in heavy meson decays such as $B \rightarrow D$
and  $B\rightarrow D^*$ \cite{Isgur90}, HQS implies that there is only
one single form factor called the Isgur-Wise function, which is
also independent of the heavy quark flavors involved.
Furthermore, the well known normalization of this universal function
at the zero-recoil point provides a model independent means of determining
the Kabayashi-Maskawa matrix element $|V_{cb}|$ from the decay data.

Formally this new symmetry
can be derived directly from quantum chromodynamics (QCD)
when $m_Q\rightarrow \infty$.
It corresponds to the symmetry limit of the so-called
heavy quark effective theory (HQET) \cite{Hill,Georgi,Luke,Mannel},
which provides a systematic $1/m_Q$ expansion of the heavy
quark sector of the QCD Lagrangian.
However, symmetry alone can only relate static and transition
properties of heavy hadrons.
In order to actually compute any physical amplitudes, such
as the universal Isgur-Wise function\cite{Isgur90}, one still has to
confront the non-perturbative QCD dynamics.
Currently, except for the lattice approach, the Isgur-Wise function
has only been computed in various hadronic models, such as the constituent
quark model \cite{Isgur91}, the bag model \cite{bag}, and QCD
sum rules \cite{Neubert92a}. It would be very interesting if one
could calculate the Isgur-Wise function, or any hadronic form factors,
directly from QCD.

Recently, there are considerable interests and activities
in the study of hadronic bound state problems using
QCD quantized on the light-front \cite{Wilson94}.  The
main advantage of this approach is that the light-front
Hamiltonian field theory provides us with a direct way of constructing
relativistic bound states by solving Schr\"{o}dinger-type
eigenstate equations in a truncated Fock space \cite{Brodsky}.
Moreover, since the boost operation on the light-front is kinematic,
it is easy to boost the resulting wave function to any other frames once the
bound state equation is solved in a particular Lorentz frame. Consequently,
once a light-front QCD description of hadronic bound states is obtained,
it is straightforward to compute
various hadronic amplitudes with arbitrary momentum transfers.

In this letter, we derive a heavy quark effective theory from QCD
quantized on the light-front. Our main purpose is to develop a formalism
which is useful towards the goal of
computing various form factors of heavy hadrons
directly from QCD.  Compared with the usual HQET in the equal-time
quantization, as we shall see later, the light-front HQET has certain
advantages that the canonical equal-time HQET does not have\cite{Suzuki91}.
Furthermore, we also find that all of the
higher order terms in the light-front HQET have rather
simple operator structures.

\vskip 0.5 true cm
\noindent {\bf 2. $1/m_Q$ Expansion of the Heavy Quark Lagrangian
on the Light-Front}

Let us begin with the QCD Lagrangian for a heavy quark:
\BE
        {\cal L} = \bar{Q} (i\! \not{\! \! D} - m_Q) Q ,
\EE
where $Q$ is the heavy quark field operator, $m_Q$ the heavy quark mass
and $D^{\mu}$ the QCD covariant derivative: $D^{\mu} = \partial^{\mu}
- ig T_a A^{\mu}_a$.
To formulate the HQET on the light-front,
the following light-front notations will be adopted:
The space-time coordinate is denoted by
$x^{\mu} = (x^+,x^-,x_{\bot})$, where $x^+ =
x^0 + x^3$ is the light-front time-like component, $x^- = x^0 - x^3$
and $x_{\bot}^i~ ( i=1,2)$ are respectively the light-front
longitudinal and transverse components.
The light-front derivatives are given by
$\partial^+=2{\partial\over\partial x^-}$,
$\partial^-=2{\partial\over\partial x^+}$, and
$\partial_{\bot}^i={\partial\over\partial x^i}$.
The product of two four-vectors is written as
$a \cdot b = {1 \over 2}( a^+b^- + a^- b^+) - a_{\bot} \cdot b_{\bot}$.

In the equal-time formalism, HQET is obtained by redefining the
heavy quark field as\cite{Georgi,Luke}:
\BE
        Q(x) = e^{-i m_Q v \cdot x} [h_v(x) +H_v(x)]   \label{rdhq1},
\EE
where $v$ is the four velocity of the heavy quark, such that $v^2=1$;
$h_v(x)$ and $H_v(x)$ are respectively
the so-called large and small components of
the heavy quark field, satisfying
$\not{\! v}h_v(x)=h_v(x)$ and $\not{\! v}H_v(x)=-H_v(x)$.
 From the QCD equation of motion, one can express $H_v(x)$ in terms of
$h_v(x)$ and show that the former is suppressed by $1/m_Q$ compared to the
later.  Using Eq.(\ref{rdhq1}) and the relation between $h_v(x)$ and
$H_v(x)$, one can systematically expand
the QCD Lagrangian in powers of $1/m_Q$,
and arrive at an effective theory for the
heavy quark\cite{Georgi,Luke,Mannel}.

In the framework of light-front quantization, the situation is
quite different.  Before taking the heavy quark mass limit,
the heavy quark field is already divided into two parts:
$Q(x) = Q_+(x) + Q_-(x)$,  with $Q_{\pm}(x) = \Lambda_{\pm} Q(x)
={1 \over 2} \gamma^0 \gamma^{\pm} Q(x)$.  The Dirac equation for $Q$
can then be rewritten as two coupled equations for $Q_\pm$:
\be
       i D^-Q_+(x) &=& ( i \alpha_{\bot} \cdot D_{\bot}
                + \beta m_Q) Q_- (x),   \label{lffd1}  \\
       i D^+Q_-(x) &=& ( i \alpha_{\bot} \cdot D_{\bot}
                + \beta m_Q) Q_+ (x),   \label{lffd2}
\ee
where $\alpha_{\bot} = \gamma^0 \gamma_{\bot}$ and $\beta = \gamma^0$.
The above equations show that only the plus-component
$Q_+(x)$ is the dynamical field.  The equation of motion for the
minus-component $Q_-(x)$ does not contain
a light-front time derivative and therefore is a light-front constraint
that determines $Q_-(x)$ from $Q_+(x)$. In terms of
$Q_+(x)$, the QCD Lagrangian (1) for the heavy quark can be rewritten as

\BE
        {\cal L} = Q_+^{\dagger} i D^- Q_+ - Q_+^{\dagger} ( i
                \alpha_{\bot} \cdot D_{\bot} + \beta m_Q ) Q_- ,
\EE
where $Q_-$ can be eliminated by Eq.(\ref{lffd2}).

To derive the light-front HQET, we use the same
redefinition of the heavy quark field as in the equal-time case,
\BE
        Q(x) = e^{-i m_Q v \cdot x} {\cal Q}_v(x),  \label{nlfq1}
\EE
but without imposing any constraint on the new variable ${\cal Q}_v$
to separate the large and small components.
It follows that
\BE
        Q_+(x) = e^{-i m_Q v \cdot x} {\cal Q}_{v+}(x) ~~,~~~
        Q_-(x) = e^{-i m_Q v \cdot x} {\cal Q}_{v-}(x) . \label{nlfq2}
\EE
Substituting these equations into Eq.(\ref{lffd2}), we obtain
\BE
        {\cal Q}_{v-} (x) = {1 \over m_Q v^+ + iD^+} \Big[i\alpha_{\bot}
                \cdot D_{\bot} + m_Q ( \alpha_{\bot} \cdot v_{\bot} +
                \beta) \Big] {\cal Q}_{v+} (x).
\EE
It is worth noting that in the ordinary light-front formulation of
field theory, the elimination of the dependent component
${Q_-}$ requires the choice of
the light-front gauge $A^+=0$, and a specification of the operator
$1/\partial^+$ which leads to severe light-front infrared problem
that has still not been completely understood\cite{Wzhang93a}.
However, for the heavy quark field with the redefinition of Eq.(\ref{nlfq1}),
the above problem does not occur since the elimination of the dependent
component ${{\cal Q}_{v-}}$ now depends on the operator $1/(m_Q
v^+ + iD^+)$ which has no infrared problem.
Moreover, it has a well defined series expansion in powers of
$iD^+/m_Q$:
\BE
        {1 \over m_Q v^+ + iD^+} = \sum_{n=1}^{\infty} \Big( {1 \over
                m_Q v^+} \Big)^n ( -iD^+)^{n-1}.
\EE
Thus,
\be
        {\cal Q}_{v-}(x) &=& \Big\{ { \alpha_{\bot} \cdot v_{\bot} + \beta
                \over v^+} + \sum_{n=1}^{\infty} \Big( {1 \over m_Q v^+}
                \Big)^{n}(-i D^+)^{n-1} ( i \vec{\alpha} \cdot \vec{D})
                \Big\} {\cal Q}_{v+}(x) \nonumber \\
        &=& \Big\{ { \alpha_{\bot} \cdot v_{\bot} + \beta \over v^+} +
                {1 \over m_Q v^+ + iD^+}(i \vec{\alpha} \cdot \vec{D})
                \Big\} {\cal Q}_{v+}(x),     \label{lfme}
\ee
where we have defined
\BE
        \vec{\alpha} \cdot \vec{D}= \alpha_{\bot} \cdot D_{\bot} -
                { \alpha_{\bot} \cdot v_{\bot} + \beta \over v^+} D^+ .
\EE

Using the Eq. (10), one can rewrite the equation of motion for
${\cal Q}_{v+}(x)$, Eq.(\ref{lffd1}), as:
\BE
        2(iv \cdot D) {\cal Q}_{v+} (x) =  (i \vec{\alpha} \cdot \vec{D})
                {v^+ \over m_Q v^+ + iD^+}(i \vec{\alpha} \cdot \vec{D})
                {\cal Q}_{v+}(x).  \label{lfhqem}
\EE
Likewise, the heavy quark QCD Lagrangian (5) can be expressed in terms
of ${\cal Q}_{v+}$ alone.
The complete $1/m_Q$ expansion is given by
\be
        {\cal L}&=& {2 \over v^+} {\cal Q}_{v+}^{\dagger} (iv \cdot D)
                {\cal Q}_{v+} - {\cal Q}_{v+}^{\dagger}(i \vec{\alpha}
                \cdot \vec{D}){1 \over m_Q v^+ + iD^+}(i \vec{\alpha}
                \cdot \vec{D}){\cal Q}_{v+}(x). \nonumber \\
        &=& {2 \over v^+} {\cal Q}_{v+}^{\dagger} (iv \cdot D)
                {\cal Q}_{v+} - \sum_{n=1}^{\infty} \Big({ 1 \over m_Q v^+}
                \Big)^n {\cal Q}_{v+}^{\dagger} \Big\{(i\vec{\alpha}
                \cdot \vec{D}) (-i D^+)^{n-1} (i \vec{\alpha} \cdot
                \vec{D}) \Big\} {\cal Q}_{v+} (x) \nonumber \\
        &=& {\cal L}_0 + \sum_{n=1}^{\infty} {\cal L}_n . \label{lfhqetl}
\ee
This is the light-front effective heavy quark Lagrangian.
It is easy to check that the equation of motion,
Eq.(\ref{lfhqem}), is consistent with this Lagrangian.  The expansion
parameter in the above Lagrangian is indeed $\Lambda_{QCD}/m_Q$ since
the operator $(-iD^+)$ picks up the ``residual'' momentum of the heavy
quark, $k^+ = p^+ - m_Q v^+$, which is of the order $\Lambda_{QCD}$.

As mentioned earlier, in the above derivation of the light-front HQET,
unlike the equal-time case,
no constraint is imposed from the start to separate the large and small
components of the heavy quark field.
In the present formalism, this separation
of large and small components is automatic.
To see this more clearly, we rewrite the above results
in covariant forms. We define
\BE
        {\cal Q}_v = {\cal Q}_{v+} + {\cal Q}_{v-} \equiv h_v^L +
                H_v^L ,
\EE
where $h_v^L$ is $m_Q$ independent and $H_v^L$ contains
all the $1/m_Q$ correction terms, viz.,
\be
        h_v^L &=& \Big\{ 1 + { \alpha_{\bot} \cdot v_{\bot} + \beta \over
                v^+} \Big\} {\cal Q}_{v+}, \label{llfhd1} \\
        H_v^L  &=& { 1 \over m_Q v^+ + iD^+}
                (i{\vec\alpha}\cdot{\vec D}) {\cal Q}_{v+} = - {\not{\! n}
\over 2(m_Q
                n \cdot v + i n \cdot D)} (i\! \not{\! \! D}) h_v^L ,
\ee
with $n^{\mu} = (0,1,0_{\bot})$.  The superscript $L$ represents the
fact that the large and small components of the heavy quark field are
separated on the light-front. One can readily prove that the
zeroth order field operator $h_v^L$ has the desired property
\BE
        \not{\! v} h_v^L = h_v^L.
\EE
Also, $H_v^L$ satisfies $\Lambda_+ H_v^L =0$;
thus all $1/m_Q$ corrections are contained
in the light-front ``bad'' component $Q_-(x)$.
This fact provides a direct connection
of heavy quark $1/m_Q$ corrections to high-twist operators\cite{Jaffe}.
In terms of $h_v^L$, the covariant form of the light-front effective
heavy quark Lagrangian reads
\be
{\cal L} &=& \overline{h}_v^L (i v \cdot D) h_v^L
            - \overline{h}_v^L (i\! \not{\! \! D}) { \not{\! n}
            \over 2(m_Q n\cdot v + i n \cdot D)} ( i\! \not{\! \! D})
            h_v^L \nonumber \\
   &=& \overline{h}_v^L (i v \cdot D) h_v^L
       - {1\over2}\sum_{l=1}^{\infty} \Big( {1 \over m_Q~n\cdot v} \Big)^l ~
       \overline{h}_v^L (i\! \not{\! \! D}) \not{\! n}
       (-in \cdot D )^{l-1}
       ( i\! \not{\! \! D}) h_v^L. \label{clfhql}
\ee

\vskip 0.5 true cm
\noindent {\bf 3. Properties of the Light-Front HQET}

In the symmetry limit, the light-front HQET reduces to
\BE
        {\cal L}_0 = {2 \over v^+} {\cal Q}_{v+}^{\dagger} (i v \cdot D)
                {\cal Q}_{v+} = \overline{h}_v^L (i v \cdot D) h_v^L ,
\EE
which clearly exhibits the flavor and spin symmetries,
since it is independent of Dirac $\gamma$-matrices and
the heavy quark mass, as in the equal-time formulation.

However, beyond the symmetry limit, the light-front HQET has several
advantages over the equal-time formulation.
In the equal-time HQET, the non-leading terms contain
high order time-derivatives; consequently
it is difficult to perform a consistent canonical
quantization beyond the limit $m_Q \rightarrow \infty$ \cite{Suzuki91}.
It is remarkable to see that in the light-front HQET, only
linear time-derivative appears, and it resides
in ${\cal L}_0$.  The presence of the matrix $ \not{\! n}$ in the
non-leading terms eliminates all light-front time derivative
terms.  This can be seen more clearly in Eq.(\ref{lfhqetl}).  Thus the
canonical quantization of light-front HQET is straightforward:
First of all, the canonical conjugate of the dynamical variable
${\cal Q}_{v+}$ is given by
\be
        \Pi_{{\cal Q}_{v+}} = { \partial {\cal L} \over \partial
                (\partial^- {\cal Q}_{v+})} = i {\cal Q}_{v+}^{\dagger},
         \label{cong}
\ee
which does not involve any $1/m_Q$ corrections. Then using the
light-front phase space quantization \cite{Wzhang93a}, we obtain the
basic anti-commutation relation:
\BE
        \{ {\cal Q}_{v+}(x)~, ~ \Pi_{{\cal Q}_{v+}}(y) \}_{x^+=y^+} =
                i\Lambda_+ \delta^3 (x-y),
\EE
which is valid to all orders in $1/m_Q$.

The second very useful property of the light-front HQET is that the
heavy quark effective Hamiltonian is well defined on the light-front.
 From Eqs.(\ref{lfhqetl}) and (\ref{cong}), we obtain the light-front
heavy quark effective Hamiltonian,
\BE
        H = \int dx^- d^2x_{\bot} {\cal H}(x)
\EE
with the Hamiltonian density ${\cal H}$ given by
\be
        {\cal H} &=& i{\cal Q}_{v+}^{\dagger} \partial^- {\cal Q}_{v+}
                - {\cal L} \nonumber \\
                &=& { 1\over iv^+} {\cal Q}^{\dagger}_{v+} (v^-\partial^+
                -2v_{\bot} \cdot \partial_{\bot} ) {\cal Q}_{v+}
                - {2g \over v^+} {\cal Q}^{\dagger}_{v+} (v \cdot A)
                {\cal Q}_{v+} + {\cal H}_{m_Q}
\ee
and
\BE
        {\cal H}_{m_Q}= \sum_{n=1}^{\infty} {\cal H}_n
                = - \sum_{n=1}^{\infty} {\cal L}_n   \label{lfhqeh}
\EE
This light-front heavy quark effective Hamiltonian can serve as
a basis for constructing heavy hadron bound states.
It is also useful for the study of heavy quark dynamics
with model light-front wave functions.
It is interesting to note that the non-leading
light-front effective Hamiltonian ${\cal H}_n$ is precisely the
minus of the corresponding effective Lagrangian ${\cal L}_n$ given
by Eq.(\ref{lfhqetl}). This simple relation is not valid in
the equal-time HQET, due to appearance of the high-order
time-derivative terms.

Furthermore, since we have not chosen any specific gauge,
and also there is no light-front infrared divergent problem for
the heavy quark sector,
short-distance QCD corrections to the heavy quark current
and the effective Lagrangian must be
the same as those calculated in the equal-time formulation\cite{Falk,Falk2}.
Of course, an explicit calculation of the short-distance effects in the
light-front HQET is needed to confirm the above statement.
We will nevertheless leave it for future investigation.

Suppose we choose the light-front gauge ($A^+=0$) in the
light-front HQET, we see immediately from Eq.(\ref{clfhql})
that, in the symmetry breaking terms,
the power of the gluon field does not increase with that of $1/m_Q$.
This property, which is unique to the light-front formulation,
may greatly simplify our treatment of $1/m_Q$ corrections.

The heavy quark current can also be systematically expanded in
$1/m_Q$ on the light-front.  For instance,
\be
    \overline{Q}^j(x)\Gamma Q^i(x)&=&e^{-i(m_{Q^j}v'-m_{Q^i}v)\cdot  x }
        {\cal Q}_{v+}^{j\dagger} \left(1 + { \alpha_{\bot} \cdot v'_{\bot}
        +\beta \over v'^+} + (-i \vec\alpha \cdot
        \stackrel{\leftarrow}{D}_{\bot} ) \sum_{n=1}^{\infty}
        \Big( {1 \over m_Qv'^+} \Big)^n (i
        \stackrel{\leftarrow}{D}^+)^{n-1} \right) \nonumber\\
   & & ~~ \times  \gamma^0 \Gamma \left(1+{\alpha_{\bot} \cdot v_{\bot}+
        \beta \over v^+} + \sum_{n=1}^{\infty} \Big( {1 \over m_Qv^+}
        \Big)^n (-iD^+)^{n-1} (i \vec\alpha \cdot \vec D_{\bot} ) \right) {\cal
        Q}^i_{v+}(x).  \label{lfcl}
\ee
In the heavy mass limit, it reduces to the following familiar from:
\BE
        \overline{Q}^j(x)\Gamma Q^i(x) = e^{-i(m_{Q^j}v'-m_{Q^i}v) \cdot x }
                \overline{h}_{v}^{jL}(x) \Gamma h_v^{iL} (x) ,
\EE
Consequences of the spin symmetry can be readily derived
using this zeroth order heavy quark current.
As an example, consider the matrix elements
\BE
        \langle P_{Q^j} (v') | \overline{h}_{v'}^{jL} \Gamma
                h_v^{iL} | P_{Q^i} (v) \rangle
                 ~~ {\rm and} ~~ \langle P^*_{Q^j} (v') |
                \overline{h}_{v'}^{jL} \Gamma h_v^{iL} | P_{Q^i} (v)
                \rangle,            \label{27}
\EE
where $\Gamma$ stands for any arbitrary gamma matrix, $P_Q$ and $P^*_Q$
represent respectively a pseudoscalar meson and a vector meson containing
a single heavy quark $Q$.
The quantum numbers of the heavy mesons can
be efficiently accounted for by the
interpolating fields\cite{Falk2,Wise}: $| P_{Q^i} (v)\rangle =
\overline{h}_v^{iL}\gamma_5 \ell_v | 0 \rangle$,
$| P^*_{Q^i} (v) \rangle =
\overline{h}_v^{iL}\not{\!\epsilon}~\ell_v | 0 \rangle$, where
$\epsilon$ is the polarization vector of the vector meson, and
$\ell_v$ represents the fully interacting light quark (or brown muck).
 From $\langle 0 | {\cal Q}_{v+} {\cal Q}^{\dagger}_{v+} | 0 \rangle = {v^+
\over 2} \Lambda_+$, it is easy to show that
\BE
        h_v^L \overline{h}_v^L = \Big( 1 + { \alpha_{\bot} \cdot v_{\bot}
                + \beta \over v^+} \Big) {v^+ \over 2} \Lambda_+ \Big( 1
                + { \alpha_{\bot} \cdot v_{\bot} + \beta \over v^+} \Big)
                \beta  = { 1+ \not{\! v} \over 2}.
\EE
Hence, in the heavy mass limit, the heavy meson decay matrix elements
on the light-front take the familiar forms:
\be
        & & \langle P_{Q^j} (v') | \overline{h}_{v'}^{jL} \Gamma
                h_v^{iL} | P_{Q^i} (v) \rangle
                = Tr\Big\{ \gamma_5 \Big( {1+ \not{\! v}'
                \over 2} \Big)\Gamma \Big( {1+\not{\! v} \over 2} \Big)
                \gamma_5 M \Big\} \\
        & & \langle P^*_{Q^j} (v') |
                \overline{h}_{v'}^{jL} \Gamma h_v^{iL} | P_{Q^i} (v)
                \rangle  =
                 Tr\Big\{ \not{\! \epsilon}^* \Big( {1 + \not{\! v}'
                \over 2} \Big) \Gamma \Big( {1+ \not{\! v} \over 2}
                \Big) \gamma_5  M \Big\}.
\ee
where M is the transition matrix element for the light quark \cite{Wise},
\BE
        M = \langle 0 | \overline{\ell}_{v'} \ell_v | 0 \rangle
                \rightarrow  \xi (v' \cdot v) I.
\EE
Thus spin symmetry implies that the transition matrix elements (\ref{27}) are
described by a single form factor $\xi(v \cdot v')$,
which is just the famous Isgur-Wise function.
Explicit calculation of the Isgur-Wise function from light-front
bound state wave function will be presented in a forthcoming paper.

\vskip 0.5 true cm
\noindent {\bf 4. Two-component Formulation of Light-Front HQET}

It is well known that, on the light-front, a Dirac field theory
can be cast in two component form if a special $\gamma$-matrix
representation is chosen\cite{Wzhang93a}:
\BE
        \gamma^0 = \left(\begin{array}{cc} 0 & -i \\
                i & 0 \end{array} \right) ~,~~
        \gamma^1 = \left(\begin{array}{cc} -i\sigma_2 & 0 \\
                0 & i \sigma_2 \end{array} \right) ~,~~
        \gamma^2 = \left(\begin{array}{cc} i\sigma_1 & 0 \\
                0 & - i\sigma_1 \end{array} \right) ~,~~
        \gamma^3 = \left(\begin{array}{cc} 0 & i \\
                i & 0 \end{array} \right) .
\EE
In this particular representation
the light-front heavy quark field has only
two components, namely,
\BE
        {\cal Q}_{v+}(x) = \left(\begin{array}{c} \xi_{v} \\
                0 \end{array} \right).
\label{33}
\EE

The physical meanings of the symmetry breaking terms
in the light-front HQET become more transparent
in this two-component formulation.
Consider the leading $1/m_Q$ correction in the light-front
effective Hamiltonian.
Without loss of generality,
we can choose a frame of reference such that the
transverse momentum of the heavy quark $v_{\bot} = 0$, and hence
$v^-=1/v^+$.
Within this frame, and adopting the light-cone gauge ($A^+=0$), the
leading order $1/m_Q$ corrections to the light-front heavy quark
effective Hamiltonian become much simpler,
\BE
        {\cal H}_1 = - {1 \over m_Q v^+} \xi_{v}^{\dagger}
                \Big\{ D_{\bot}^2  + {1 \over v^{+2}} (\partial^+)^2
                + g \sigma_3 B^3 - {g \over v^+}
                (\sigma_1E^2-\sigma_2E^1)
                \Big\} \xi_{v} ,
\EE
where $B^3 ={1\over ig} [iD_{\bot}^1 ,iD_{\bot}^2]$ is the chromomagnetic
field along the $z$-direction, and $E^i=\partial^+ A_{\bot}^i$ is the
transverse component of the chromoelectric field.
The physical meaning of each term in ${\cal H}_1$ is very clear.
The first two terms represent correction to the kinetic energy
of the heavy quark;
the third and the fourth terms are spin (helicity) dependent
coupling to the gluon field.
Likewise, the $1/m_Q$ corrections to the heavy quark currents in the
two-component formulation can also be easily obtained from
Eq.(\ref{lfcl}) and Eq.(\ref{33}).


\vskip 0.5 true cm
\noindent{\bf 5. Summary}

In this letter, we have formulated the HQET on the light front.
Consequences of the
HQS as derived in the equal-time formalism
can readily be reproduced here.  In addition, canonical quantization
of light-front HQET is very simple, in contrast to the equal-time
HQET where the quantization procedure becomes non-trivial due to
the occurrence of higher order time derivatives.  Also, the
interaction part of the light-front effective Hamiltonian is simply
the minus of the light-front interaction Lagrangian.  Such a relation
does not exist in the equal-time HQET.
Finally we stress that the main purpose of this letter is to provide
a basic formalism for constructing light-front heavy hadron bound
states from QCD.  The results obtained in this work are also useful
in studying heavy quark dynamics with model light-front wave functions.
All practical applications will however be left to
future publications.

\acknowledgements

We thank Tung-Mow Yan for suggesting this investigation.
This work is supported in part by National Science Council of
the Republic of China under Grant Nos. NSC 84-2816-M-001-012L,
NSC 84-2112-M-009-024, and NSC 84-2112-M-001-036.

\newpage


\end{document}